\documentclass[twocolumn,showpacs,preprintnumbers,amsmath,amssymb]{revtex4}
\usepackage{graphicx}
\begin{document}
\title{Photon exchange and correlations transfer in atom-atom entanglement dynamics}
\author{Juan Le\'on}
\email{leon@imaff.cfmac.csic.es} \homepage{http://www.imaff.csic.es/pcc/QUINFOG/}
\author{Carlos Sab\'in}
\email{csl@imaff.cfmac.csic.es} \homepage{http://www.imaff.csic.es/pcc/QUINFOG/} \affiliation{ Instituto de F\'{i}sica
Fundamental, CSIC
 \\
Serrano 113-B, 28006 Madrid, Spain.\\}
\date{}

\begin{abstract}
We analyze the entanglement dynamics of a system composed by a pair of neutral two-level atoms that are initially entangled, and
the electromagnetic field, initially in the vacuum state, within the formalism of perturbative quantum field theory up to the
second order. We show that entanglement sudden death and revival can occur while the atoms remain spacelike-separated and
therefore cannot be related with photon exchange between the atoms. We interpret these phenomena as the consequence of a
transfer of atom-atom entanglement to atom-field entanglement and viceversa. We also consider the different bi-partitions of the
system, finding similar relationships between their entanglement evolutions.
\end{abstract}
\maketitle

\section{Introduction}

Entanglement between qubits may disappear in a finite time when the qubits interact with a reservoir. This is commonly known as
``entanglement sudden death'' (ESD). After its discovery \cite{zyczk,diosi,yueberly}, the phenomenon has attracted great
attention (for instance, \cite{yueberlyII,Yonac,jamroz,ficektanasI,compagnoII,paz,lastrasolano,ficek,cole}) and has been
observed experimentally \cite{mafalda}.

ESD shows up in a variety of systems that can be roughly divided in two sets: those in which the qubits interact individually
with different reservoirs and those in which they interact with a common environment. In particular, in
\cite{jamroz,ficektanasI,ficektanasII} a system of a pair of two-level atoms interacting with a common electromagnetic vacuum is
considered. The dynamics of the system is given in all the cases by the Lehmberg-Agarwal master equation \cite{lehmberg,
agarwal} which is derived with the rotating wave approximation (RWA) and the Born-Markov approximation. Recently, non-Markovian
\cite{compagnoII} and non-RWA \cite{treschinos} effects have been considered in systems of qubits coupled individually to
different reservoirs. There are good reasons for going beyond the Markovian and RWA scenario in the case of a pair of two-level
atoms in the electromagnetic vacuum. For short enough times non-RWA contributions are relevant \cite{conjuanII} and a proper
analysis of causality issues can only be performed if they are taken into account \cite{powerthiru,milonni,compagnoI}. Besides,
as we shall show in this paper the death of the entanglement between the atoms is related with the birth of entanglement between
the atoms and the field, and therefore the field is actually a non- Markovian reservoir. This was also the case in
\cite{Yonac,lastrasolano,ficek} with different reservoirs.

In \cite{conjuan,conjuanII,conjuanIII} we have applied the formalism of perturbative quantum electrodynamics (QED) to the system
of a pair of neutral two-level atoms interacting locally with the electromagnetic field, and for initially separable states
analyzed the generation of entanglement. This is a non-Markovian, non-RWA approach. The use of the Lehmberg-Agarwal master
equation can be seen as a coarse-grained in time approximation to the perturbative treatment \cite{cohentannoudji}. The first
goal of this paper is to apply also the QED formalism to analyze the ESD in these systems for initially entangled atomic states,
comparing the results with the previously obtained \cite{jamroz,ficektanasI} with master equations. We will focus mainly on the
range $r/(c\,t)\approx1$, $r$ being the interatomic distance and $t$ the interaction time, in order to investigate the role of
locality. We will also consider for the first time in these systems the rest of pairwise concurrences, namely the entanglement
of each atom with the field, and multipartite entanglement, following the spirit of \cite{Yonac,lastrasolano,ficek,cole}. While
the mentioned papers deal with a four qubit model, our model here consists in two qubits (the atoms) and a qutrit (the
electromagnetic field, which may have 0, 1 or 2 photons). We shall show that the phenomenon of revival of entanglement after the
ESD \cite{ficektanasI} can occur for $r>c\,t$, and therefore is not related with photon exchange as is usually believed. We will
see that atom-atom disentaglement is connected with the growth of atom-field entanglement and viceversa. Similar relationship
will be obtained among the ``atom-(atom+field)'' and ``field-(atom+atom)'' entanglements.

The reminder of the paper is organized as follows. In section II we will describe the Hamiltonian and the time evolution from
the initial state of the system. In section III we will obtain the reduced state of the atoms and analyze the behavior of its
entanglement. In section IV the same will be performed with the reduced state of each atom and the field, comparing the
entanglement cycle with the one obtained in the previous section. Tripartite entanglement will be considered in section V in
terms of the entanglement of all the different bi-partitions of the system, and we conclude in section VI with a summary of our
results.

\section{Hamiltonian and state evolution}

To address the atom-field interactions, we assume that the relevant wavelengths and the interatomic separation are much larger
than the atomic dimensions.  The dipole approximation, appropriate to these conditions, permits the splitting of the system
Hamiltonian into two parts $H = H_0 + H_I$ that are separately gauge invariant. The first part is the Hamiltonian in the absence
of interactions other than the potentials that keep $A$ and $B$ stable, $H_0 = H_A + H_B + H_{\mbox{field}}$. The second
contains all the interaction of the atoms with the field, which in the dipole approximation we will use is given by:
\begin{equation}
H_I = - \frac{1}{\epsilon_0}\sum_{n=A,B} \mathbf{d}_n(\mathbf{x}_n,t)\,\mathbf{D}(\mathbf{x}_n,t),\label{a}
\end{equation}
where $\mathbf{D}$ is the electric displacement field, and $\mathbf{d}_n \,=\,\sum_i\, e\,\int d^3 \mathbf{x}_i\,
\langle\,E\,|\,(\mathbf{x}_i-\mathbf{x}_n)\,|\,G\,\rangle$ is the electric dipole moment of atom $n$, that we will take of equal
magnitude for both atoms $(\mathbf{d}=\mathbf{d_A}=\mathbf{d_B})$ as required by angular momentum conservation in the photon
exchange. $|\,E\,\rangle$ and $|\,G\,\rangle$ will denote the excited and ground states of the atoms, respectively.

In what follows we choose a system  given initially by an atomic entangled state, with the field in the vacuum state
$|\,0\rangle$:
\begin{equation}
|\,\psi\,\rangle_0\,=(\,\alpha\, |\,E\,E\,\rangle +\,\beta\,|\,G\,G\,\rangle) \cdot|\,0\,\rangle.\label{b}
\end{equation}
The system then evolves under the effect of the interaction during a lapse of time $t$ into a state:
\begin{equation}
|\,\psi\,\rangle_t = T (e^{-i\, \int_0^t\,dt'\, H_I\,(t')/\hbar})\,|\,\psi\,\rangle_0, \label{c}
\end{equation}
$T$ being the time ordering operator. Up to second order in perturbation theory, (\ref{c}) can be given in the interaction
picture as
\begin{eqnarray}
|\mbox{atom} 1,\mbox{atom} 2,\mbox{field}\rangle_{t} = \,\alpha\,  |\,E\,E\,0\,\rangle_t
 +\,\beta\,|\,G\,G\,0\,\rangle_t  \label{d}
\end{eqnarray}
where
\begin{eqnarray}
|\,E\,E\,0\,\rangle_t=\,((1+a)\,|\,E\,E\rangle + b\,|\,G\,G\rangle)\,|\,0\rangle\nonumber\\
 +(u_A\,|\,G\,E\,\rangle+ u_B\,|\,E\,G\,\rangle)\,|\,1\,\rangle+
(f\,|\,E\,E\rangle+ g\,|\,G\,G\rangle)\,|\,2\rangle \label{e}
\end{eqnarray}
and
\begin{eqnarray}
|\,G\,G\,0\rangle_t=\,((1+a')\,|\,G\,G\rangle + b'\,|\,E\,E\rangle)\,|\,0\rangle\nonumber\\
 +(v_A\,|\,E\,G\,\rangle+ v_B\,|\,G\,E\,\rangle)\,|\,1\,\rangle+
(f'\,|\,G\,G\rangle+ g'\,|\,E\,E\rangle)\,|\,2\rangle \label{f}
\end{eqnarray}
where
\begin{eqnarray}
a&=&\frac{1}{2}\,\theta(t_1-t_2)\langle\,0|\mathcal{S}_A^+\,(t_1)
\mathcal{S}_A^-\,(t_2)+\mathcal{S}_B^+\,(t_1)\mathcal{S}_B^-\,(t_2)|0\rangle\nonumber\\
a'&=&\frac{1}{2}\,\theta(t_1-t_2)\langle\,0|\mathcal{S}_A^-\,(t_1)
\mathcal{S}_A^+\,(t_2)+\mathcal{S}_B^-\,(t_1)\mathcal{S}_B^+\,(t_2))|0\rangle\nonumber\\
b&=&\langle\,0|T(\mathcal{S}^-_B\, \mathcal{S}^-_A)\,|0\rangle, \,b'=\langle\,0|T(\mathcal{S}^+_B\,
\mathcal{S}^+_A)\,|0\rangle,\nonumber\\
u_A\,&=&\,\langle\,1\,|\, \mathcal{S}^-_A\,|\,0\,\rangle,\, v_A\,=\,\langle\,1\,|\,
\mathcal{S}^+_A\,|\,0\,\rangle \label{g}\\
u_B\,&=&\,\langle\,1\,|\, \mathcal{S}^-_B\,|\,0\,\rangle,\, v_B\,=\,\langle\,1\,|\, \mathcal{S}^+_B\,|\,0\,\rangle\nonumber\\
f&=&\frac{1}{2}\,\theta(t_1-t_2)\langle\,2|\mathcal{S}_A^+\,(t_1)
\mathcal{S}_A^-\,(t_2)+\mathcal{S}_B^+\,(t_1)\mathcal{S}_B^-\,(t_2)|0\rangle\nonumber\\
f'&=&\frac{1}{2}\,\theta(t_1-t_2)\langle\,2|\mathcal{S}_A^-\,(t_1)
\mathcal{S}_A^+\,(t_2)+\mathcal{S}_B^-\,(t_1)\mathcal{S}_B^+\,(t_2))|0\rangle,\nonumber\\
g&=&\langle\,2|T(\mathcal{S}^-_B\, \mathcal{S}^-_A\,)|0\rangle,\,g'=\langle\,2|T(\mathcal{S}^+_B\,
\mathcal{S}^+_A\,)|0\rangle\nonumber
\end{eqnarray}
being $\mathcal{S}\,=\,- \frac{i}{\hbar}  \int_0^t\, dt\, H_{I}(t')= S^+\,+\, S^-$ , $T$ the time ordering operator and
$|\,n\,\rangle,\,\, n=\,0,\,1,\,2$ is a shorthand for the state of $n$ photons with definite momenta and polarizations, i.e.
$|\,1\,\rangle\,=\,|\mathbf{k},\, \mathbf{\epsilon}\,\rangle$, etc. Here, $a$ and $a'$ describe intra-atomic radiative
corrections, $u_A\, (u_B)$ and $v_A\, (v_B)$ single photon emission by  atom $A$ ($B$), and $g$ and $g'$ by both atoms, while
$f$ and $f'$ correspond to two photon emission by a single atom.  Only $b$ and $b'$ correspond to interaction between both
atoms. The sign of the superscripts is associated to the energy difference between the initial and final atomic states of each
emission or absorption. In Quantum Optics, virtual terms like $a'$, $b$, $b'$, $v_A$, $v_B$, $f$, $f'$ and $g'$, which do not
conserve energy and appear only at very short times, are usually neglected by the introduction of a RWA. In the dipole
approximation the actions $\hbar\, \mathcal{S}^{\pm}$ in (\ref{e}) reduce to
\begin{eqnarray}
\mathcal{S}^{\pm}\,=\, \frac{i}{\hbar}  \int_0^t\, dt' \: e^{\pm i\Omega t'}\, \mathbf{d}\,\mathbf{E}(\mathbf{x},t')\label{h}
\end{eqnarray}
where  $\Omega = \omega_E -\omega_G$ is the transition frequency, and we are neglecting atomic recoil. This depends on the
atomic properties $\Omega$ and $\mathbf{d}$, and on the interaction time $t$. In our calculations we will take $(\Omega
|\mathbf{d}|/e c) = 5\,\cdot 10^{-3}$, which is of the same order as the 1s $\rightarrow$ 2p transition in the hydrogen atom,
consider $\Omega\,t \gtrsim 1$, and focus mainly on the cases $(r/c\,t)\simeq 1$. Therefore,  $|\,E\,\rangle$ is actually a
triply degenerate state $|\,E\,,m\rangle$ with $m=0,\pm1$ and we will average over two different independent possibilities for
dipole orientations: $\mathbf{d}_A=\mathbf{d}_B=\mathbf{d}=d\,\mathbf{u}_z$ for transitions with $\Delta m=0$ \cite{milonniII}
and $\mathbf{d}=d\,(\mathbf{u}_x \pm i \mathbf{u}_y)/\sqrt{2}$ \cite{milonniII} for transitions with $\Delta m=\pm1$.

\section{Sudden death and revival of atom-atom entanglement}

After tracing over all the states of the field, the density matrix of the atomic state $\rho_{AB}$ takes the form (in the basis
$\{|EE\rangle,|EG\rangle,|GE\rangle,|GG\rangle\}$):
\begin{eqnarray}
\rho_{AB}=\frac{1}{N}\left( \begin{array}{c c c c}\rho_{11}&0&0&\rho_{14} \\0&\rho_{22}&\rho_{23}&0\\0&\rho_{23}^*&\rho_{33}&0\\
\rho_{14}^*&0&0&\rho_{44}\end{array}\right)\label{i}
\end{eqnarray}
where
\begin{eqnarray}
\rho_{11}&=&|\,\alpha\,(1\,+\,a)\,+\,\beta\,b'\,|^2\,+\,|\,\alpha\,f\,+\,\beta\,g'\,|^2,\nonumber\\
\rho_{22}&=&\rho_{33}=|\,\alpha\,|^2\,|\,u\,|^2\,+\,|\,\beta\,|^2\,|\,v\,|^2+2\,Re\,(\alpha\,\beta^*\,l^*)\nonumber\\
\rho_{44} &=&|\,\alpha\,b\,+\,\beta\,(1\,+\,a')\,|^2\,+\,|\,\alpha\,g\,+\,\beta\,f'\,|^2,\label{j}\\
\rho_{14} &=& |\,\alpha\,|^2\,((1\,+\,a)\,b*\,+\,f\,g^*)+|\,\beta\,|^2\,((1\,+\,a')^*\,b'\nonumber\\
&+&\,g'\,f'^*)+\alpha\,\beta^*((1+a)\,(1+a')\,+\,f\,f'^*)\nonumber\\
&+& \beta\,\alpha^*(b'\,b^*+\,g'\,g^*)\nonumber\\
\rho_{23} &=& |\,\alpha\,|^2\,u_B\,u_A^*+|\,\beta\,|^2\,v_A\,v_B^*\,+\,2\,Re\,(\alpha\,\beta^*\,u\,v^*)\nonumber\\
N&=&\,\rho_{11}\,+\,\rho_{22}\,+\,\rho_{33}+\,\rho_{44}\nonumber
\end{eqnarray}
where $|\,u\,|^2\,=\,|\,u_A\,|^2\,=\,|\,u_B\,|^2$, $|\,v\,|^2\,=\,|\,v_A\,|^2\,=\,|\,v_B\,|^2$,
$l\,=\,u_A\,v_B^*\,=\,u_B\,v_A^*$ and $u\,v^*\,=\,u_A^*\,v_A^*=\,u_B\,v_B^*$.

The computation of $a$, $b$, etc. can be performed following the lines given in the Appendix A of \cite{conjuan}, where they
were computed for the initial state $|\,E\,G\,\rangle$ and only for $\Delta m=0$. In terms of $z\,=\,\Omega\,r/c$ and
$x\,=\,r/c\,t$, being $r$ the interatomic distance, we find:
\begin{eqnarray}
a&=&\frac{4\,i\,K\,z^3}{3\,x}\,(\ln{(1-\frac{z_{max}}{z})}\,+\,i\,\pi),\nonumber\\
a'&=&\frac{-4\,i\,K\,z^3}{3\,x}\,\ln{(1+\frac{z_{max}}{z})}\nonumber\\
b&=&=b'^*=\frac{\alpha\,d_i\,d_j}{\pi\,e^2}(-\mathbf{\nabla}^2\delta_{ij}+\nabla_i\nabla_j)\,I, \label{k}
\end{eqnarray}
with $K=\alpha\,|\,\mathbf{d}\,|^2/(e^2\,r^2)$ and $I=I_+\,+\,I_-$, where:
\begin{eqnarray}
I_{\pm}&=&\frac{-i\,e^{-i\frac{z}{x}}}{2\,z}\,[\,\pm\,2\cos(\,\frac{z}{x}\,)\,e^{\pm\,i\,z}\,Ei(\mp\,i\,z)
+\,e^{-i\,z\,(1\pm\frac{1}{x})}\nonumber\\&\ &
Ei(i\,z\,(1\pm\frac{1}{x}))\,-\,e^{i\,z\,(1\pm\frac{1}{x})}\,Ei(-i\,z\,(1\pm\frac{1}{x}))\,]\label{l}
\end{eqnarray}
for $x>1$, having the additional term $-2\,\pi\,i\,e^{i\,z\,(1-1/x)}$ otherwise.

$|\,u\,|^2$, $|\,v\,|^2$, $l$, $u_B\,u_A^*$, $v_A\,v_B^*$ and $u\,v^*$ have been computed in \cite{conjuan}. Besides:
\begin{eqnarray}
g&=&u_B\,u'_A\,+\,u_A\,u'_B\, , g'=v_A\,v'_B\,+\,v_B\,v'_A\nonumber\\
f&=&\theta(t_1-t_2)(\,v_A\,(t_1)\,u'_A\,(t_2)\,+\,u_A\,(t_1)\,v'_A\,(t_2)\nonumber\\
&+&\,v_B\,(t_1)\,u'_B\,(t_2)\,+\,u_B\,(t_1)\,v'_B\,(t_2)\,)\label{m}\\
f'&=&\theta(t_1-t_2)(\,u_A\,(t_1)\,v'_A\,(t_2)\,+\,u'_A\,(t_1)\,v_A\,(t_2)\nonumber\\
&+&\,u_B\,(t_1)\,v'_B\,(t_2)\,+\,u'_B\,(t_1)\,v_B\,(t_2)\,)\nonumber
\end{eqnarray}
where the primes are introduced to discriminate between the two single photons.

We will use the concurrence $\mathbb{C}(\rho)$ \cite{wootters} to compute the entanglement, which for a state like (\ref{i}) is
given by, if $\,\sqrt{\rho_{22}\,\rho_{33}}+\,|\,\rho_{23}\,|\,>\,\sqrt{\rho_{11}\,\rho_{44}}\,+\,|\,\rho_{14}\,|$
\begin{equation}
\mathbb{C}(\rho_{AB})=\mbox{max}\left(\frac{2\,(\,|\,\rho_{23}|\,-\sqrt{\rho_{11}\,\rho_{44}})}{N}\,,\,0\right)\label{n}
\end{equation}
and
\begin{equation}
\mathbb{C}(\rho_{AB})=\mbox{max}\left(\frac{2\,(\,|\,\rho_{14}|\, -\sqrt{\rho_{22}\,\rho_{33}})}{N}\,,\,0\right)\label{o}
\end{equation}
otherwise.

If we take $\alpha=\sqrt{p}$ and $\beta=\sqrt{1-p}$, we find that ESD appears at a range of values of $p$ that decreases with
increasing $r$, in agreement with \cite{ficektanasI}. Although this would suggest that ESD disappear for $r$ large enough, we
find that there are high values of $p$ for which ESD exists for arbitrary large $r$. In Fig. 1, we represent
$\mathbb{C}(\rho_{AB})$ in front of $x$ for different values of $z$ and $p=0.98$. ESD occurs at $z/x=\Omega\,t$ of the order of
$10^7$. Thus, as $z$ (that is $r$) grows, ESD is shifted to higher values of $x$. It is also interesting to analyze the
phenomenon of entanglement revival, discovered in these systems in \cite{ficektanasI}. We find that the dark periods
\cite{ficektanasI}  between death and revival has larger time durations for increasing $z$. Besides, although in
\cite{ficektanasI} the revival is described as a consequence of the photon exchange, for $r$ sufficiently large both the ESD and
the revival can occur for $x>1$, where photon exchange is not allowed. We think that the explanation for entanglement revival is
closer to the spirit of \cite{Yonac,ficek} where entanglement revival between noninteracting atoms is interpreted as coming from
entanglement transfer between different parts of the system. We shall discuss this point in the following sections.
\begin{figure}[h]
\includegraphics[width=0.45\textwidth]{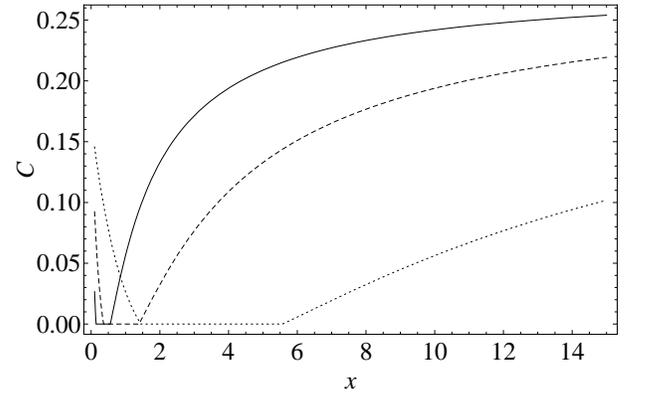}
\caption{Concurrence $\mathbb{C}(\rho_{AB})$ in front of $x=r/c\,t$ for $p=0.98$ and $z=\Omega r/c= 2\cdot10^6$ (solid line),
$5\cdot10^6$ (dashed line) and $2\cdot10^7$ (dotted line). In the latter case sudden death and revival of entanglement occur for
$x>1$.}
\end{figure}

In Fig. 2 we sketch the dependence with $p$. Although sudden death and revivals appear in a very restricted range of the
parameter, they are only a particular case of the generic behavior of entanglement observed in a wider range, which can be
described as disentanglement up to a minimum value and growth of quantum correlations since then.

\begin{figure}[h]
\includegraphics[width=0.45\textwidth]{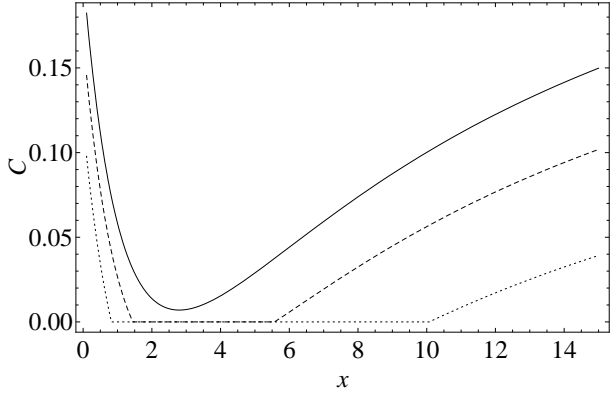}
\caption{Concurrence  $\mathbb{C}(\rho_{AB})$ in front of $x=r/c\,t$ for $z=\Omega r/c= 2\cdot10^7$  and $p=0.97$  (solid line),
$p=0.98$ (dashed line) and $p=0.99$ (dotted line). In the first case, entanglement decreases as $t$ grows up to a minimum value
and begin to grow since then. This behavior becomes entanglement sudden death and revival when the minimum value is 0 for higher
values of $p$. $\mathbb{C}(\rho_{AB})$ tends to 0 as $x\rightarrow\infty$ and $p\rightarrow1$.}
\end{figure}

\section{Atom-field entanglement}

Tracing (\ref{d}) over states of atom A (B) the reduced atom-field density matrix $\rho_{BF}$ ($\rho_{AF}$) is obtained. Taking
the basis $\{|\,E\,0\,\rangle,|\,E\,1\,\rangle,|\,E\,2\,\rangle,|\,G\,0\,\rangle,|\,G\,1\,\rangle,|\,G\,2\,\rangle\}$, we have:
\begin{eqnarray}
\rho_{BF}=\rho_{AF}=\frac{1}{N'}\left( \begin{array}{c c c c c c}\rho'_{11}&0&\rho'_{13}&0&\rho'_{15}&0
\\0&\rho'_{22}&0&\rho'_{24}&0&\rho'_{26}
\\ \rho'^{*}_{13}&0&\rho'_{33}&0&\rho'_{35}&0\\
0&\rho'^{*}_{24}&0&\rho'_{44}&0&\rho'_{46}\\
\rho'^{*}_{15}&0&\rho'^{*}_{35}&0&\rho'_{55}&0\\0&\rho'^{*}_{26}&0&\rho'^{*}_{46}&0&\rho'_{66}
\end{array}\right)\label{p}
\end{eqnarray}
with
\begin{eqnarray}
\rho'_{11}&=&|\,\alpha\,(1\,+\,a)\,+\,\beta\,b'\,|^2,\,\rho'_{22}=\rho'_{55}=\rho_{22}\nonumber\\
\rho'_{33}&=&|\,\alpha\,f\,+\,\beta\,g'\,|^2,\,\rho'_{44}=|\,\alpha\,b\,+\,\beta\,(1\,+\,a')\,|^2\nonumber\\
\rho'_{66}&=&|\,\alpha\,g\,+\,\beta\,f'\,|^2,\,\rho'_{13}=(\alpha\,(1\,+\,a)+\beta\,b')\,(\alpha\,f\,+\,\beta\,g')\nonumber\\
\rho'_{15}&=&(\alpha\,(1\,+\,a)+\beta\,b')\,(\alpha\,u_{B}\,+\,\beta\,v_A)^* \label{q}\\
\rho'_{24}&=&(\alpha\,u_{A}\,+\,\beta\,v_B)\,(\beta\,(1+a')\,+\,\alpha\,b)^*\nonumber\\
\rho'_{26}&=&(\alpha\,u_{A}\,+\,\beta\,v_B)\,(\alpha\,g\,+\,\beta\,f')^*\nonumber\\
\rho'_{35}&=& (\alpha\,f\,+\,\beta\,g')\,(\alpha\,u_{B}\,+\,\beta\,v_{A})^*\nonumber\\
\rho'_{46}&=& (\alpha\,b\,+\,\beta\,(1\,+a'))\,(\alpha\,g\,+\,\beta\,f')^*\nonumber\\
N'&=&\,\rho'_{11}\,+\,\rho'_{22}\,+\,\rho'_{33}+\,\rho'_{44}+\,\rho'_{55}+\,\rho'_{66}\nonumber
\end{eqnarray}
There are no operational generalizations of concurrence for mixed states in $2\times3$ dimensions like the ones in Eq.
(\ref{p}). We will use the negativity \cite{vidalwerner} $\mathbb{N}(\rho)$, which is the absolute value of the sum of the
negative eigenvalues of the partial transposes of a state $\rho$. For the $2\times2$ and $2\times3$ cases $\mathbb{N} (\rho)>0$
is a necessary and sufficient condition for $\rho$ to be entangled.

Up to second order in perturbation theory, we have that $N'=N$ and that the nonzero eigenvalues of the partial transposes of
both $\rho_{BF}$ and $\rho_{AF}$ are
\begin{equation}
\lambda_{\pm}=\frac{\rho'_{11}+\rho'_{55}\pm\sqrt{(\rho'_{11}-\rho'_{55})^2+4|\rho'_{24}|^2}}{2\, N'}\label{r}
\end{equation}
and
\begin{equation}
\lambda'_{\pm}=\frac{\rho'_{44}+\rho'_{55}\pm\sqrt{(\rho'_{44}-\rho'_{22})^2+4|\rho'_{15}|^2}}{2\, N'}\label{s}
\end{equation}
being zero the other two. In Eqs. (\ref{r}) and (\ref{s}) only the terms up to second order are retained.  Therefore, if
$|\rho'_{24}|^2>\rho'_{11}\,\rho'_{55}$ then $\lambda_{-}<0$ and if $|\rho'_{15}|^2>\rho'_{22}\,\rho'_{44}$ then
$\lambda'_{-}<0$.

In Fig. 3 we represent $\mathbb{N}(\rho_{BF})=\mathbb{N}(\rho_{AF})$ in front of $x$ for same values of $p$ and $z$ of Fig. 1.
We see that the negativity grows from 0 at $x\rightarrow\infty$ ($t=0$) to its maximum value and then starts to decrease and
eventually vanishes, following the opposite cycle to the entanglement of $\rho_{AB}$.
\begin{figure}[h]
\includegraphics[width=0.45\textwidth]{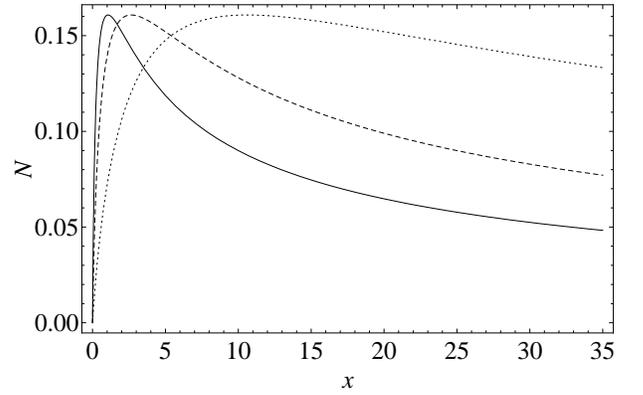}
\caption{Negativity  $\mathbb{N}(\rho_{BF})=\mathbb{N}(\rho_{AF})$ in front of $x=r/c\,t$ for $p=0.98$ and $z=\Omega r/c=
2\cdot10^6$ (solid line), $5\cdot10^6$ (dashed line) and $2\cdot10^7$ (dotted line). Entanglement increases from 0 at
$x\rightarrow\infty$ up to a maximum value and then decreases and vanishes eventually.}
\end{figure}

Although it would be interesting to look for conservation rules of entanglement like the ones in \cite{Yonac,ficek,cole}, this
search is beyond the focus of this paper since in our study we are using different entanglement measures in Hilbert spaces of
different dimensions. Besides, except for the concurrence between atoms $A$ and $B$, the rest of the concurrences in the
mentioned papers have not obvious counterparts in our case. But it is clear that in general the entanglement cycle between atoms
is correlated with the entanglement cycle between each atom and the field, as can be seen in Fig. 4 in a particular case.
Although atom-field entanglement may change while the other remains zero, both entanglements cannot increase or decrease at the
same time.
\begin{figure}[h]
\includegraphics[width=0.45\textwidth]{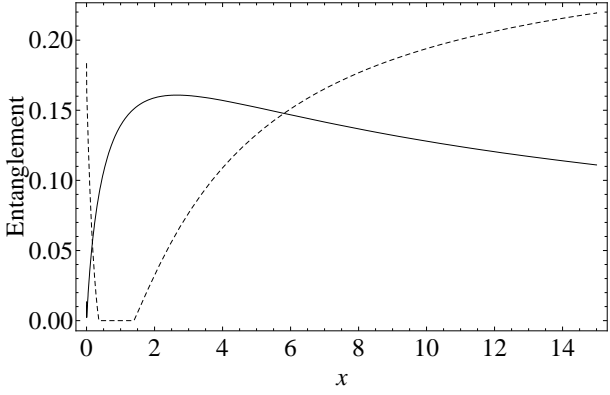}
\caption{Negativity $\mathbb{N}(\rho_{BF})=\mathbb{N}(\rho_{AF})$ (solid line) and concurrence $\mathbb{C}(\rho_{AB})$ (dashed
line) in front of $x=r/c\,t$ for $p=0.98$ and $z=\Omega r/c= 5\cdot10^6$. Entanglement atom-atom cycle is clearly correlated
with the atom-field cycle, although the sum is not a conserved quantity. Although atom-field entanglement may change while the
other remains zero, both entanglements cannot increase or decrease at the same time.}
\end{figure}

\section{Tripartite entanglement}

Tripartite entanglement has been widely studied in terms of the entanglement of the different bipartitions $A-BC$, $B-AC$,
$C-AB$ in the system \cite{shan,conguillermo,pascazio}, where $A$, $B$ and $C$ stand for the three parties. Here, we will
compute the I concurrences \cite{rungta} $\mathbb{C}_{A-BF}$, $\mathbb{C}_{B-AF}$, $\mathbb{C}_{F-AB}$, where
$\mathbb{C}_{J-KL}=\sqrt{2(1-Tr\, \rho_{J}^2)}$, where $J$ runs form $A$ to $F$ and $KL$ from $BF$ to $AB$ respectively, being
$\rho_{J}$ the reduced density matrix of $J$. $A$ and $B$ stand for the atoms, and $F$ for the field.

\begin{figure}[h]
\includegraphics[width=0.45\textwidth]{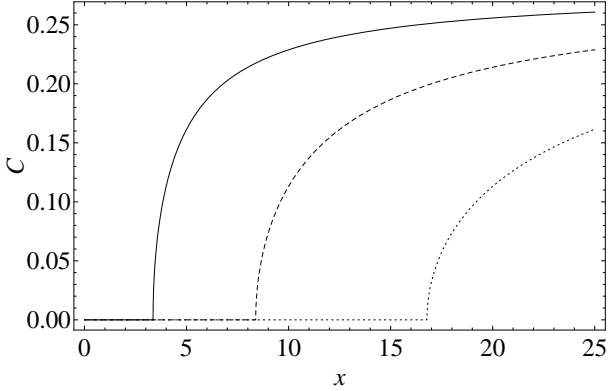}
\caption{I concurrence $\mathbb{C}_{A-BF}=\mathbb{C}_{B-AF}$  in front of $x=r/c\,t$ for $p=0.98$ and $z=\Omega r/c=
2\cdot10^5$(solid line), $5\cdot10^5$ (dashed line) and $1\cdot10^6$ (dotted line). Entanglement disappears faster than the
entanglement between the atoms (Fig.1) and remains 0 since then.}
\end{figure}

\begin{figure}[h]
\includegraphics[width=0.45\textwidth]{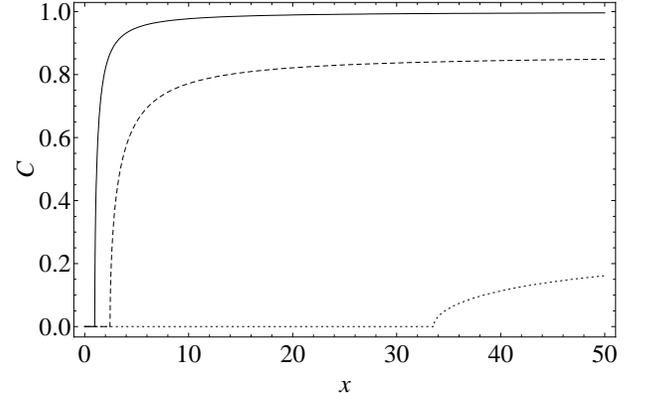}
\caption{I concurrence $\mathbb{C}_{A-BF}=\mathbb{C}_{B-AF}$  in front of $x=r/c\,t$ for $z=\Omega r/c= 2\cdot10^6$ and $p=0.50$
(solid line), $p=0.75$ (dashed line) and $p=0.98$ (dotted line). Entanglement sudden death occurs for a wider range than the
entanglement between the atoms (Fig.2).}
\end{figure}

Tracing (\ref{d}) over $BF$ $(AF)$, we find the following density matrices $\rho_{A}$ $(\rho_{B})$:
\begin{eqnarray}
\rho_{A}=\rho_{B}=\frac{1}{N_{A}}\left( \begin{array}{c c }\rho_{A11}&0
\\0&\rho_{A22}\end{array}\right)\label{t}
\end{eqnarray}
where $\rho_{A11}=\rho'_{11}+\rho'_{33}+\rho_{22}$ and $\rho_{A22}=\rho'_{44}+\rho'_{66}+\rho_{22}$ and
$N_{A}=\rho_{A11}+\rho_{A22}$. In Fig. 5 we sketch the behavior of $\mathbb{C}_{A-BF}$ and $\mathbb{C}_{B-AF}$ in front of $x$
for different values of $z$. Entanglement vanishes before the death of the entanglement between $A$ and $B$, and does not have a
revival. Besides, ESD appears in a wider range of $p$, as can be seen in Fig. 6.

\begin{figure}[h]
\includegraphics[width=0.45\textwidth]{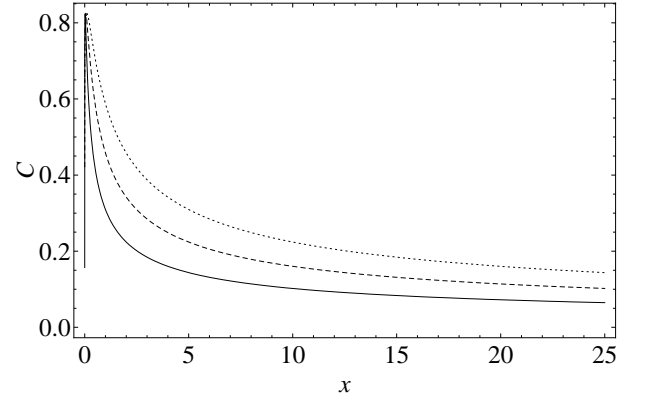}
\caption{I concurrence $\mathbb{C}_{F-AB}$  in front of $x=r/c\,t$ for $p=0.98$ and $z=\Omega r/c= 2\cdot10^5$ (solid line),
$5\cdot10^5$ (dashed line) and $1\cdot10^6$ (dotted line). Entanglement grows from 0 to its maximum value at $x\approx0.1$ and
then decreases.}
\end{figure}

\begin{figure}[h]
\includegraphics[width=0.45\textwidth]{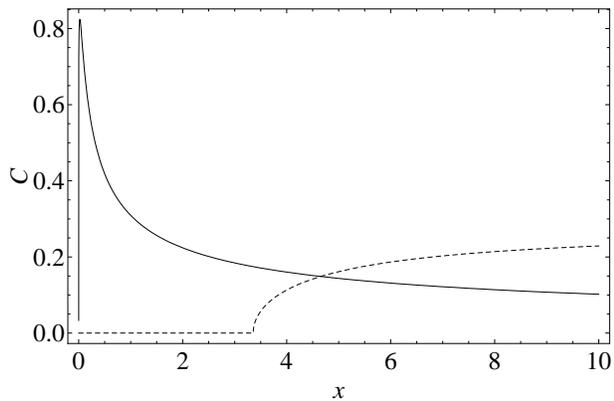}
\caption{I concurrence $\mathbb{C}_{F-AB}$ (solid line) and $\mathbb{C}_{A-BF}= \mathbb{C}_{B-AF}$ (dashed line) in front of
$x=r/c\,t$ for $z=\Omega r/c= 2\cdot10^5$ and $p=0.98$. Both magnitudes cannot increase or decrease at the same time.}
\end{figure}
Now, tracing (\ref{d}) over $AB$ we obtain the reduced density matrix of the field $\rho_{F}$:
\begin{eqnarray}
\rho_{F}=\frac{1}{N_{F}}\left( \begin{array}{c c c}\rho_{F11}&0&\rho_{F13}
\\0&\rho_{F22}&0\\\rho^*_{F13}&0&\rho_{F33}\end{array}\right)\label{u}
\end{eqnarray}
where $\rho_{F11}=\rho'_{11}+\rho'_{44}$, $\rho_{F22}=2\rho_{22}$, $\rho_{F33}=\rho'_{33}+\rho'_{66}$,
$\rho_{F02}=\rho'_{13}+\rho'_{46}$ and $N_{F}=\rho_{F11}+\rho_{F22}+\rho_{F33}$. In Fig. 7 we represent $\mathbb{C}_{F-AB}$ in
front of $x$ for the same values of $z$ and $p$ as in Fig. 5. Entanglement grows from 0 to a maximum value at $x\approx0.1$ and
then decreases. The growth of $\mathbb{C}_{F-AB}$ is correlated with the decrease of $\mathbb{C}_{A-BF}$ and $\mathbb{C}_{B-AF}$
in the same way as the magnitudes analyzed in the previous section, as can be seen in Fig. 8 for a particular case.

\section{Conclusions}
We have analyzed in a previously unexplored spacetime region the entanglement dynamics of a system consisting in a pair of
neutral two-level atoms $A$ and $B$ interacting with a common electromagnetic field $F$. At $t=0$ atoms are in the Bell state
$\sqrt{p}\,|\,E\,E\,\rangle\,+\,\sqrt{1-p}\,|\,G\,G\,\rangle$ and the field in the vacuum state. The evolution of this state has
been considered within the non-Markovian, non-RWA approach of quantum electrodynamics up to second order in perturbation theory.
We find ESD and revival of entanglement in the reduced state of the atoms, in a range of $p$ that decreases with the interatomic
distance $r$, in agreement with the results obtained with master equations \cite{ficektanasI}. For $r$ large enough, we find
that the revival of entanglement can occur with $r>c\,t$ and therefore is not a consequence of photon exchange between the
atoms. We find that this phenomenon is strongly related to the transfer of entanglement between the different subsystems of two
parties that coexist in the entire system: we obtain sort of entanglement cycle for the atom-field reduced states opposite to
the atom-atom one. We have considered also the different bi-partitions of the system, namely $A-BF$, $B-AF$ and $F-AB$, finding
similar relationships between their entanglement cycles.
\begin{acknowledgments}
This work was supported by Spanish MEC FIS2005-05304 and CSIC 2004 5 OE 271 projects.
\end{acknowledgments}


\begin{thebibliography}{31}
\expandafter\ifx\csname natexlab\endcsname\relax\def\natexlab#1{#1}\fi \expandafter\ifx\csname bibnamefont\endcsname\relax
  \def\bibnamefont#1{#1}\fi
\expandafter\ifx\csname bibfnamefont\endcsname\relax
  \def\bibfnamefont#1{#1}\fi
\expandafter\ifx\csname citenamefont\endcsname\relax
  \def\citenamefont#1{#1}\fi
\expandafter\ifx\csname url\endcsname\relax
  \def\url#1{\texttt{#1}}\fi
\expandafter\ifx\csname urlprefix\endcsname\relax\def\urlprefix{URL }\fi \providecommand{\bibinfo}[2]{#2}
\providecommand{\eprint}[2][]{\url{#2}}

\bibitem[{\citenamefont{\ifmmode~\dot{Z}\else \.{Z}\fi{}yczkowski
  et~al.}(2001)\citenamefont{\ifmmode~\dot{Z}\else \.{Z}\fi{}yczkowski,
  Horodecki, Horodecki, and Horodecki}}]{zyczk}
\bibinfo{author}{\bibfnamefont{K.}~\bibnamefont{\ifmmode~\dot{Z}\else
  \.{Z}\fi{}yczkowski}},
  \bibinfo{author}{\bibfnamefont{P.}~\bibnamefont{Horodecki}},
  \bibinfo{author}{\bibfnamefont{M.}~\bibnamefont{Horodecki}},
  \bibnamefont{and}
  \bibinfo{author}{\bibfnamefont{R.}~\bibnamefont{Horodecki}},
  \bibinfo{journal}{Phys. Rev. A} \textbf{\bibinfo{volume}{65}},
  \bibinfo{pages}{012101} (\bibinfo{year}{2001}).

\bibitem[{\citenamefont{Di{\'o}si}(2003)}]{diosi}
\bibinfo{author}{\bibfnamefont{L.}~\bibnamefont{Di{\'o}si}},
  \bibinfo{journal}{Lect. Notes Phys.} \textbf{\bibinfo{volume}{622}},
  \bibinfo{pages}{15} (\bibinfo{year}{2003}).

\bibitem[{\citenamefont{Yu and Eberly}(2004)}]{yueberly}
\bibinfo{author}{\bibfnamefont{T.}~\bibnamefont{Yu}} \bibnamefont{and}
  \bibinfo{author}{\bibfnamefont{J.~H.} \bibnamefont{Eberly}},
  \bibinfo{journal}{Phys. Rev. Lett.} \textbf{\bibinfo{volume}{93}},
  \bibinfo{pages}{140404} (\bibinfo{year}{2004}).

\bibitem[{\citenamefont{Yu and Eberly}(2006)}]{yueberlyII}
\bibinfo{author}{\bibfnamefont{T.}~\bibnamefont{Yu}} \bibnamefont{and}
  \bibinfo{author}{\bibfnamefont{J.~H.} \bibnamefont{Eberly}},
  \bibinfo{journal}{Phys. Rev. Lett.} \textbf{\bibinfo{volume}{97}},
  \bibinfo{pages}{140403} (\bibinfo{year}{2006}).

\bibitem[{\citenamefont{Y\"{o}na\c{c} et~al.}(2007)\citenamefont{Y\"{o}na\c{c},
  Yu, and Eberly}}]{Yonac}
\bibinfo{author}{\bibfnamefont{M.}~\bibnamefont{Y\"{o}na\c{c}}},
  \bibinfo{author}{\bibfnamefont{T.}~\bibnamefont{Yu}}, \bibnamefont{and}
  \bibinfo{author}{\bibfnamefont{J.~H.} \bibnamefont{Eberly}},
  \bibinfo{journal}{J. Phys. B: At. Mol. Opt. Phys.}
  \textbf{\bibinfo{volume}{40}}, \bibinfo{pages}{S45} (\bibinfo{year}{2007}).

\bibitem[{\citenamefont{Jamr{\'o}z}(2006)}]{jamroz}
\bibinfo{author}{\bibfnamefont{A.}~\bibnamefont{Jamr{\'o}z}},
  \bibinfo{journal}{J. Phys. A: Math. Gen.} \textbf{\bibinfo{volume}{39}},
  \bibinfo{pages}{7727} (\bibinfo{year}{2006}).

\bibitem[{\citenamefont{Ficek and Tana{\'s}}(2006)}]{ficektanasI}
\bibinfo{author}{\bibfnamefont{Z.}~\bibnamefont{Ficek}} \bibnamefont{and}
  \bibinfo{author}{\bibfnamefont{R.}~\bibnamefont{Tana{\'s}}},
  \bibinfo{journal}{Phys. Rev. A} \textbf{\bibinfo{volume}{74}},
  \bibinfo{pages}{024304} (\bibinfo{year}{2006}).

\bibitem[{\citenamefont{Bellomo et~al.}(2008)\citenamefont{Bellomo, Franco, and
  Compagno}}]{compagnoII}
\bibinfo{author}{\bibfnamefont{B.}~\bibnamefont{Bellomo}},
  \bibinfo{author}{\bibfnamefont{R.~L.} \bibnamefont{Franco}},
  \bibnamefont{and} \bibinfo{author}{\bibfnamefont{G.}~\bibnamefont{Compagno}},
  \bibinfo{journal}{Phys. Rev. A} \textbf{\bibinfo{volume}{77}},
  \bibinfo{pages}{032342} (\bibinfo{year}{2008}).

\bibitem[{\citenamefont{Paz and Roncaglia}(2008)}]{paz}
\bibinfo{author}{\bibfnamefont{J.~P.} \bibnamefont{Paz}} \bibnamefont{and}
  \bibinfo{author}{\bibfnamefont{A.~J.} \bibnamefont{Roncaglia}},
  \bibinfo{journal}{Phys. Rev. Lett.} \textbf{\bibinfo{volume}{100}},
  \bibinfo{pages}{220401} (\bibinfo{year}{2008}).

\bibitem[{\citenamefont{L{\'o}pez et~al.}(2008)\citenamefont{L{\'o}pez, Romero,
  Lastra, Solano, and Retamal}}]{lastrasolano}
\bibinfo{author}{\bibfnamefont{C.~E.} \bibnamefont{L{\'o}pez}},
  \bibinfo{author}{\bibfnamefont{G.}~\bibnamefont{Romero}},
  \bibinfo{author}{\bibfnamefont{F.}~\bibnamefont{Lastra}},
  \bibinfo{author}{\bibfnamefont{E.}~\bibnamefont{Solano}}, \bibnamefont{and}
  \bibinfo{author}{\bibfnamefont{J.~C.} \bibnamefont{Retamal}},
  \bibinfo{journal}{Phys. Rev. Lett.} \textbf{\bibinfo{volume}{101}},
  \bibinfo{pages}{080503} (\bibinfo{year}{2008}).

\bibitem[{\citenamefont{Chan et~al.}()\citenamefont{Chan, Reid, and
  Ficek}}]{ficek}
\bibinfo{author}{\bibfnamefont{S.}~\bibnamefont{Chan}},
  \bibinfo{author}{\bibfnamefont{M.~D.} \bibnamefont{Reid}}, \bibnamefont{and}
  \bibinfo{author}{\bibfnamefont{Z.}~\bibnamefont{Ficek}},
  \eprint{arXiv[quant-ph]:0810.3050}.

\bibitem[{\citenamefont{Cole}()}]{cole}
\bibinfo{author}{\bibfnamefont{J.~H.} \bibnamefont{Cole}},
  \eprint{arXiv[quant-ph]:0809.1764}.

\bibitem[{\citenamefont{Almeida et~al.}(2007)\citenamefont{Almeida, de~Melo,
  Hor-Meyll, Salles, Walborn, Ribeiro, and Davidovich}}]{mafalda}
\bibinfo{author}{\bibfnamefont{M.~P.} \bibnamefont{Almeida}},
  \bibinfo{author}{\bibfnamefont{F.}~\bibnamefont{de~Melo}},
  \bibinfo{author}{\bibfnamefont{M.}~\bibnamefont{Hor-Meyll}},
  \bibinfo{author}{\bibfnamefont{A.}~\bibnamefont{Salles}},
  \bibinfo{author}{\bibfnamefont{S.~P.} \bibnamefont{Walborn}},
  \bibinfo{author}{\bibfnamefont{P.~H.~S.} \bibnamefont{Ribeiro}},
  \bibnamefont{and}
  \bibinfo{author}{\bibfnamefont{L.}~\bibnamefont{Davidovich}},
  \bibinfo{journal}{Science} \textbf{\bibinfo{volume}{316}},
  \bibinfo{pages}{579} (\bibinfo{year}{2007}).

\bibitem[{\citenamefont{Ficek and Tana{\'s}}(2008)}]{ficektanasII}
\bibinfo{author}{\bibfnamefont{Z.}~\bibnamefont{Ficek}} \bibnamefont{and}
  \bibinfo{author}{\bibfnamefont{R.}~\bibnamefont{Tana{\'s}}},
  \bibinfo{journal}{Phys. Rev. A} \textbf{\bibinfo{volume}{77}},
  \bibinfo{pages}{054301} (\bibinfo{year}{2008}).

\bibitem[{\citenamefont{Lehmberg}(1970)}]{lehmberg}
\bibinfo{author}{\bibfnamefont{R.~H.} \bibnamefont{Lehmberg}},
  \bibinfo{journal}{Phys. Rev. A} \textbf{\bibinfo{volume}{2}},
  \bibinfo{pages}{883} (\bibinfo{year}{1970}).

\bibitem[{\citenamefont{Agarwal}(1974)}]{agarwal}
\bibinfo{author}{\bibfnamefont{G.~S.} \bibnamefont{Agarwal}},
  \bibinfo{journal}{Springer Tracts Mod. Phys} \textbf{\bibinfo{volume}{70}},
  \bibinfo{pages}{1} (\bibinfo{year}{1974}).

\bibitem[{\citenamefont{Wang et~al.}()\citenamefont{Wang, Zhang, and
  Liang}}]{treschinos}
\bibinfo{author}{\bibfnamefont{F.-Q.} \bibnamefont{Wang}},
  \bibinfo{author}{\bibfnamefont{Z.-M.} \bibnamefont{Zhang}}, \bibnamefont{and}
  \bibinfo{author}{\bibfnamefont{R.-S.} \bibnamefont{Liang}},
  \eprint{arXiv[quant-ph]:0806.3306}.

\bibitem[{\citenamefont{Le{\'o}n and Sab{\'i}n}(2008)}]{conjuanII}
\bibinfo{author}{\bibfnamefont{J.}~\bibnamefont{Le{\'o}n}} \bibnamefont{and}
  \bibinfo{author}{\bibfnamefont{C.}~\bibnamefont{Sab{\'i}n}},
  \bibinfo{journal}{Phys. Rev. A} \textbf{\bibinfo{volume}{78}},
  \bibinfo{pages}{052314} (\bibinfo{year}{2008}).

\bibitem[{\citenamefont{Power and Thirunamachandran}(1997)}]{powerthiru}
\bibinfo{author}{\bibfnamefont{E.~A.} \bibnamefont{Power}} \bibnamefont{and}
  \bibinfo{author}{\bibfnamefont{T.}~\bibnamefont{Thirunamachandran}},
  \bibinfo{journal}{Phys. Rev. A} \textbf{\bibinfo{volume}{56}},
  \bibinfo{pages}{3395} (\bibinfo{year}{1997}).

\bibitem[{\citenamefont{Milonni et~al.}(1995)\citenamefont{Milonni, James, and
  Fearn}}]{milonni}
\bibinfo{author}{\bibfnamefont{P.~W.} \bibnamefont{Milonni}},
  \bibinfo{author}{\bibfnamefont{D.~F.~V.} \bibnamefont{James}},
  \bibnamefont{and} \bibinfo{author}{\bibfnamefont{H.}~\bibnamefont{Fearn}},
  \bibinfo{journal}{Phys. Rev. A} \textbf{\bibinfo{volume}{52}},
  \bibinfo{pages}{1525} (\bibinfo{year}{1995}).

\bibitem[{\citenamefont{Biswas et~al.}(1990)\citenamefont{Biswas, Compagno,
  Palma, Passante, and Persico}}]{compagnoI}
\bibinfo{author}{\bibfnamefont{A.~K.} \bibnamefont{Biswas}},
  \bibinfo{author}{\bibfnamefont{G.}~\bibnamefont{Compagno}},
  \bibinfo{author}{\bibfnamefont{G.~M.} \bibnamefont{Palma}},
  \bibinfo{author}{\bibfnamefont{R.}~\bibnamefont{Passante}}, \bibnamefont{and}
  \bibinfo{author}{\bibfnamefont{F.}~\bibnamefont{Persico}},
  \bibinfo{journal}{Phys. Rev. A} \textbf{\bibinfo{volume}{42}},
  \bibinfo{pages}{4291} (\bibinfo{year}{1990}).

\bibitem[{\citenamefont{Le{\'o}n and Sab{\'i}n}({\natexlab{a}})}]{conjuan}
\bibinfo{author}{\bibfnamefont{J.}~\bibnamefont{Le{\'o}n}} \bibnamefont{and}
  \bibinfo{author}{\bibfnamefont{C.}~\bibnamefont{Sab{\'i}n}},
  \eprint{arXiv[quant-ph]:0804.4641}.

\bibitem[{\citenamefont{Le{\'o}n and Sab{\'i}n}({\natexlab{b}})}]{conjuanIII}
\bibinfo{author}{\bibfnamefont{J.}~\bibnamefont{Le{\'o}n}} \bibnamefont{and}
  \bibinfo{author}{\bibfnamefont{C.}~\bibnamefont{Sab{\'i}n}},
  \eprint{arXiv[quant-ph]:0805.2110}.

\bibitem[{\citenamefont{Cohen-Tannoudji
  et~al.}(1998)\citenamefont{Cohen-Tannoudji, Dupont-Roc, and
  Grynberg}}]{cohentannoudji}
\bibinfo{author}{\bibfnamefont{C.}~\bibnamefont{Cohen-Tannoudji}},
  \bibinfo{author}{\bibfnamefont{J.}~\bibnamefont{Dupont-Roc}},
  \bibnamefont{and} \bibinfo{author}{\bibfnamefont{G.}~\bibnamefont{Grynberg}},
  \emph{\bibinfo{title}{Atom-photon interactions}} (\bibinfo{publisher}{Wiley
  Interscience}, \bibinfo{address}{New York}, \bibinfo{year}{1998}).

\bibitem[{\citenamefont{Milonni and Knight}(1974)}]{milonniII}
\bibinfo{author}{\bibfnamefont{P.~W.} \bibnamefont{Milonni}} \bibnamefont{and}
  \bibinfo{author}{\bibfnamefont{P.~L.} \bibnamefont{Knight}},
  \bibinfo{journal}{Phys. Rev. A} \textbf{\bibinfo{volume}{10}},
  \bibinfo{pages}{1096} (\bibinfo{year}{1974}).

\bibitem[{\citenamefont{Hill and Wootters}(1997)}]{wootters}
\bibinfo{author}{\bibfnamefont{S.}~\bibnamefont{Hill}} \bibnamefont{and}
  \bibinfo{author}{\bibfnamefont{W.~K.} \bibnamefont{Wootters}},
  \bibinfo{journal}{Phys.\ Rev.\ Lett.} \textbf{\bibinfo{volume}{78}},
  \bibinfo{pages}{5022} (\bibinfo{year}{1997}).

\bibitem[{\citenamefont{{Vidal} and {Werner}}(2002)}]{vidalwerner}
\bibinfo{author}{\bibfnamefont{G.}~\bibnamefont{{Vidal}}} \bibnamefont{and}
  \bibinfo{author}{\bibfnamefont{R.~F.} \bibnamefont{{Werner}}},
  \bibinfo{journal}{\pra} \textbf{\bibinfo{volume}{65}},
  \bibinfo{pages}{032314} (\bibinfo{year}{2002}).

\bibitem[{\citenamefont{{Song} and {Yu}}(2004)}]{shan}
\bibinfo{author}{\bibfnamefont{H.-S.} \bibnamefont{{Song}}} \bibnamefont{and}
  \bibinfo{author}{\bibfnamefont{C.-s.} \bibnamefont{{Yu}}},
  \bibinfo{journal}{Phys. Lett. A} \textbf{\bibinfo{volume}{330}},
  \bibinfo{pages}{377} (\bibinfo{year}{2004}).

\bibitem[{\citenamefont{{C. Sab\'{\i}n } and {G. Garc\'{\i}a-Alcaine
  }}(2008)}]{conguillermo}
\bibinfo{author}{\bibnamefont{{C. Sab\'{\i}n }}} \bibnamefont{and}
  \bibinfo{author}{\bibnamefont{{G. Garc\'{\i}a-Alcaine }}},
  \bibinfo{journal}{Eur. Phys. J. D} \textbf{\bibinfo{volume}{48}},
  \bibinfo{pages}{435} (\bibinfo{year}{2008}).

\bibitem[{\citenamefont{Facchi et~al.}(2006)\citenamefont{Facchi, Florio, and
  Pascazio}}]{pascazio}
\bibinfo{author}{\bibfnamefont{P.}~\bibnamefont{Facchi}},
  \bibinfo{author}{\bibfnamefont{G.}~\bibnamefont{Florio}}, \bibnamefont{and}
  \bibinfo{author}{\bibfnamefont{S.}~\bibnamefont{Pascazio}},
  \bibinfo{journal}{Phys. Rev. A} \textbf{\bibinfo{volume}{74}},
  \bibinfo{pages}{042331} (\bibinfo{year}{2006}).

\bibitem[{\citenamefont{Rungta et~al.}(2001)\citenamefont{Rungta,
  Bu\ifmmode~\check{z}\else \v{z}\fi{}ek, Caves, Hillery, and
  Milburn}}]{rungta}
\bibinfo{author}{\bibfnamefont{P.}~\bibnamefont{Rungta}},
  \bibinfo{author}{\bibfnamefont{V.}~\bibnamefont{Bu\ifmmode~\check{z}\else
  \v{z}\fi{}ek}}, \bibinfo{author}{\bibfnamefont{C.~M.} \bibnamefont{Caves}},
  \bibinfo{author}{\bibfnamefont{M.}~\bibnamefont{Hillery}}, \bibnamefont{and}
  \bibinfo{author}{\bibfnamefont{G.~J.} \bibnamefont{Milburn}},
  \bibinfo{journal}{Phys. Rev. A} \textbf{\bibinfo{volume}{64}},
  \bibinfo{pages}{042315} (\bibinfo{year}{2001}).

\end{thebibliography}
\end{document}